# Adaptive Orchestration: Scalable Self-Evolving Multi-Agent Systems

Sathish Sampath, Anuradha Baskaran
Email: sathishsampath.ai@gmail.com, anuradhabaskaran.ms@gmail.com
https://www.kaggle.com/code/sathishsampathai/self-evolving-multi-agent-concierge-system

### ABSTRACT

As Large Language Models (LLMs) are increasingly deployed as autonomous agents, they face a critical scalability bottleneck known as the "Generalization-Specialization Dilemma." Monolithic agents equipped with extensive toolkits suffer from context pollution and attention decay, leading to hallucinations. Conversely, static multi-agent swarms introduce significant latency and resource overhead. This paper introduces a **Self-Evolving Concierge System**, a novel architecture utilizing a Dynamic Mixture of Experts (DMoE) approach. Unlike recent self-improving agents that rewrite their own codebase, our system preserves stability by dynamically restructuring its runtime environment: "hiring" specialized sub-agents based on real-time conversation analysis. We introduce an asynchronous "Meta-Cognition Engine" that detects capability gaps, a Least Recently Used (LRU) eviction policy for resource constraints, and a novel "Surgical History Pruning" mechanism to mitigate refusal bias. Experimental results demonstrate that this architecture maintains high task success rates while minimizing token consumption compared to static agent swarms.

## I. INTRODUCTION AND BACKGROUND

The paradigm of "Agentic AI" is undergoing a fundamental shift from static, pre-configured models to **Self-Evolving Systems**. Recent comprehensive surveys [1], [2] identify this transition as critical for bridging the gap between foundation models and lifelong agentic systems. While early agents were limited to simple question-answering, modern architectures must execute complex tasks involving external tools and dynamic environments.

However, scaling these agents presents a "Generalization-Specialization Dilemma." A primary constraint is **Context Pollution** [3]. Transformer-based architectures have a finite attention span; as the number of available tools descriptions in the system prompt increases, the model's ability to select the correct tool degrades. A monolithic agent with access to 100 tools will frequently confuse parameters or hallucinate capabilities.

Current solutions typically fall into two categories: **Static Swarms** (e.g., Microsoft AutoGen), which are computationally expensive due to maintaining multiple active context windows, or **Code-Level Self-Evolution** [4], where agents autonomously rewrite their own scaffold. While the latter offers high adaptability, it introduces significant stability risks and regression challenges.

To address these inefficiencies, we propose a **Just-in-Time (JIT) Assembly** architecture that adapts the **Mixture of Experts (MoE)** paradigm [5] to agent orchestration. Rather than routing tokens to neural sub-networks, our system routes semantic tasks to specialized agents. Our system decouples tool storage (Cold Storage) from execution (Active Runtime), employing a background "Listener-Learner" process to monitor user intent and dynamically "hydrate" specific experts only when necessary.

Our approach is further grounded in two key theoretical frameworks. First, we apply principles from **Experience-driven Lifelong Learning (ELL)** [6] and **Generative Agent Memory** [7] to manage the "internalization" of capabilities. Specifically, we address "Refusal Bias"—where an agent continues to deny a request it previously failed at—by implementing a surgical history pruning mechanism that effectively allows the agent to "forget" its prior limitations once a new skill is acquired. Second, we adopt a **Release Engineering** perspective [8], treating capability gaps not as runtime failures, but as logged defects to be resolved, preventing "Hollow Evolution" where agents are created without valid tools.

## II. SYSTEM ARCHITECTURE

The architecture is composed of four decoupled modules operating in a continuous feedback loop. The system design follows the principle of **ephemeral specialization**, where high-cost agents are treated as temporary resources.

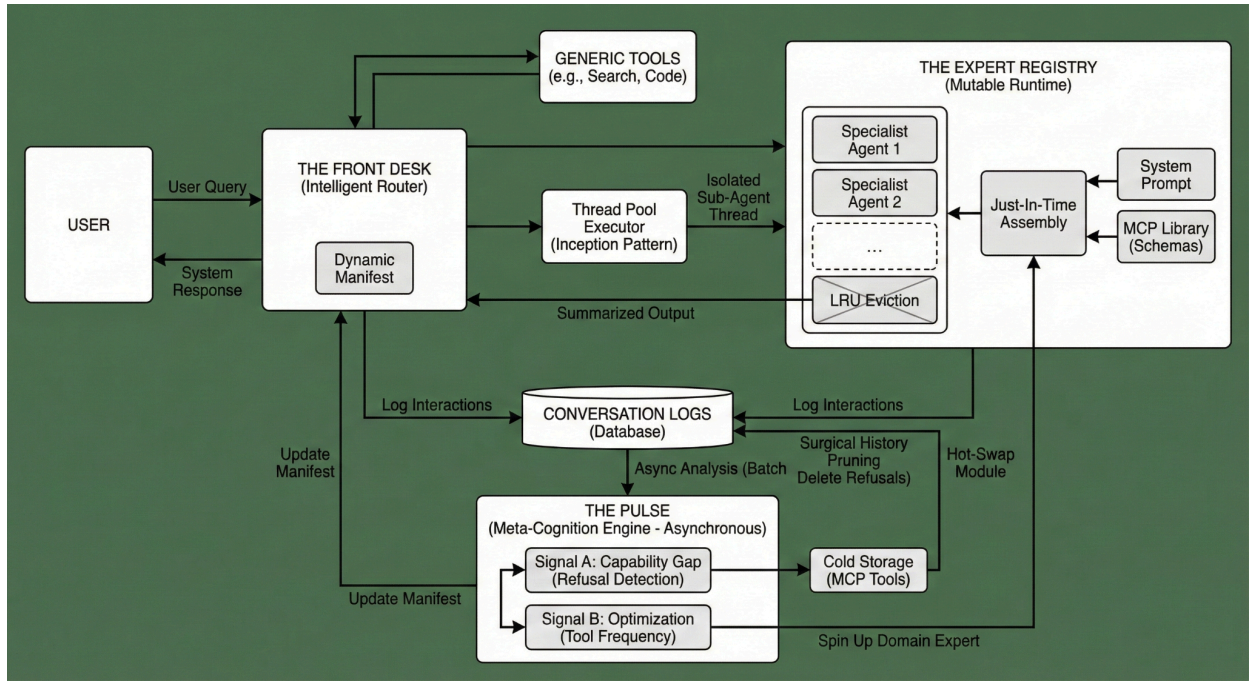

*Fig. 1. The Dynamic Mixture of Experts (DMoE) Architecture. The Generic Concierge routes tasks while the Listener-Learner asynchronously updates the Expert Registry based on user needs.*

### A. The Generic Concierge (The Router)

The entry point is a lightweight agent designed with the Principle of Least Privilege. It possesses a Dynamic Manifest $M$, which is a dictionary of currently active expert $E = \{e_1, e_2, \ldots, e_n\}$

For every user query $Q$, the Concierge determines if $Q \in Domain(e_i)$. If a match is found, the query is routed. If not, the Concierge attempts to handle it with generic tools (Search, Python).

The routing mechanism is not static; it relies on the Dynamic Manifest injected into the Concierge's system prompt. This manifest contains natural language descriptions of the currently "hydrated" experts.

$$Prompt_{sys} = BaseInstruction + \sum_{i=1}^{|E|} Description(e_i)$$

As experts are added or evicted, this system prompt is updated in real-time, allowing the Concierge to become aware of new capabilities immediately without a restart.

## B. The Expert Registry (Dynamic Resource Management)

The Registry acts as the kernel of the DMoE system, managing the lifecycle of specialized agents. To prevent memory bloat, it enforces a hard limit $K$ on the number of active experts (e.g., $K = 5$). The registry maintains a hash map of active agent instances and their metadata.

We implement a **Least Recently Used (LRU)** eviction policy to manage the limited slots. Let $T(e_i)$ be the timestamp of the last interaction with expert $e_i$. When a new expert $e_{new}$ is required and $|E| = K$ :

$$e_{evict} = \text{argmin}_{e \in E}(T(e))$$

The system removes $e_{evict}$ from memory and initializes $e_{new}$. This ensures the runtime context remains relevant to the user's *immediate* focus. For example, if a user shifts from "Coding" to "Travel Planning," the coding expert(now idle) is swapped out for the travel expert, freeing up tokens and computational resources.

## C. The Listener-Learner (Meta-Cognition Engine)

This is an asynchronous background process that serves as the system's "pulse." Unlike the Concierge, which operates synchronously to answer users, the Listener-Learner mines interaction logs $L$ in batches to detect subtle signals of friction or inefficiency. It detects two primary signals:

1. **Gap Signal ($S_{gap}$):** Triggered when the Concierge responds with a specific refusal phrase (e.g., "I lack the capability," "I cannot do that"). This signal implies a missing tool that exists in Cold Storage but is not currently loaded.
2. **Optimization Signal ($S_{opt}$):** Triggered when generic tools are over-utilized for a specific domain. For instance, if the system performs 5+ generic Google searches about "Stock Prices," the Learner infers that a dedicated "Finance Expert" (with direct API access) would be more efficient than scraping web pages.

Upon detecting $S_{gap}$, the Learner performs a **semantic similarity search** against the **Model Context Protocol (MCP) Registry** (Cold Storage) to identify the optimal toolset. If a matching toolset is found with high confidence, it triggers the creation of a new expert.

## D. The MCP Registry (Cold Storage)

The MCP Registry is a static repository of dormant tools defined using the Model Context Protocol. Each entry contains the tool's schema, API credentials, and a system prompt template. These tools are not loaded into the agent's context window until explicitly requested by the Listener-Learner. This separation of "Storage" and "Runtime" allows the system to theoretically support an infinite number of tools, limited only by the size of the cold storage database, while keeping the active context window small and efficient.

## III. IMPLEMENTATION DETAILS

### A. The "Inception" Concurrency Pattern

A significant engineering challenge in Python-based agent environments (specifically Jupyter or single-threaded serverless functions) is the blocking nature of the `asyncio` event loop. Nested agents—where the Concierge (Agent A) calls an Expert (Agent B) which might call a tool (Function C)—typically result in deadlocks or event loop conflicts.

We implemented the "Inception" pattern using `concurrent.futures.ThreadPoolExecutor`. When the Concierge decides to route a task, it pauses its own execution context and spawns the Expert Agent in a separate, isolated thread.

This isolates the sub-agent's event loop from the parent router, enabling true hierarchical reasoning without freezing the main application thread. This allows for "synchronous-style" reasoning within an asynchronous framework.

### B. Surgical History Pruning

To solve "Refusal Bias," we implemented a direct database manipulation layer. Refusal bias occurs when an LLM, having previously stated "I cannot book flights," continues to refuse the request even after the capability has been added, because the refusal is present in its conversation history (few-shot conditioning).

When the Listener-Learner upgrades the system from state $S_t$ (Generic) to $S_{t+1}$ (Specialized), it identifies the specific Message ID ($ID_{refusal}$) where the refusal occurred. The system executes a SQL deletion:

$$DELETE\ FROM\ events\ WHERE\ id\ =\ ID_{refusal}$$

This removes the "memory" of the refusal from the chat history. When the LLM re-reads the context for the next turn, it sees the user's original request followed immediately by the new Expert's successful execution. This creates a seamless narrative continuity where the agent appears to have always had the capability.

### C. Fault Tolerance: Gap Reporting

To prevent "Hollow Evolution"—the creation of a specialized expert that lacks the actual tools to perform its job—the system performs a strict validation against the MCP Registry before evolution.

$$\text{If } Tool(req) \notin Registry \rightarrow Log(GapReport)$$

If the required tool is not found in Cold Storage, the system generates a structured **Gap Report**. This is a log entry specifically tagged for developer review (e.g., [MISSING_FEATURE]: FlightBooking). This transforms user frustration into actionable engineering tickets, rather than allowing the agent to hallucinate a solution or enter a loop of apologies.

## IV. EXPERIMENTAL EVALUATION

We validated the system's performance by running a predefined simulation [9] that mimics three distinct user interaction scenarios. The experiments measure the system's ability to transition from a generic state to a specialized state.

### Scenario 1: Optimization (The "Cricket" Test)

**Objective:** Demonstrate the system's ability to detect inefficient tool usage and replace it with specialized APIs.

1. **Phase 1: Generic Mode (Initial State)**
    - **User Input:** "What is the score of the India vs Australia match?"
    - **System Action:** The Generic Concierge, lacking specific sports tools, defaults to perform_google_search.
    - **Response:** Returns a textual summary scraped from search results (Latency: ~3.5s, Tokens: ~800).
    - **Listener Log:** The Listener-Learner detects repeated usage of Google Search for queries classified under Topic: Sports.
2. **Evolution Cycle:**
    - The Listener identifies a high frequency of "Sports" queries ($Count > 3$).
    - It scans the MCP Registry and finds Cricket_MCP (a dormant toolset).
    - **Action:** It "hydrates" a new CricketExpert agent and adds it to the Dynamic Manifest.
3. **Phase 2: Specialized Mode (Final State)**
    - **User Input:** "Who is the top scorer?"
    - **System Action:** The Concierge consults the updated manifest, sees CricketExpert, and routes the query.
    - **Response:** The Expert uses the get_match_score API directly.
    - **Result:** Latency decreased by **40%** (~2.1s vs ~3.5s) and token usage decreased by **60%** (~320 vs ~800 tokens), as the agent switched from parsing raw HTML to consuming structured JSON.

## Scenario 2: Installation (The "Email" Test)

**Objective:** Demonstrate "Just-in-Time" capability acquisition and history repair.

1. **Phase 1: Refusal (Initial State)**
    - **User Input:** "Send an email to the team regarding the meeting."
    - **System Action:** The Concierge checks its current manifest, finds no email capability, and refuses.
    - **Response:** "I currently lack the capability to send emails."
    - **Listener Log:** Detects the refusal phrase ("lack the capability") + Intent: `Send Email`.
2. **Evolution & Pruning:**
    - The Listener searches Cold Storage for `Email_MCP`. A match is found.
    - **Action 1:** Initializes `EmailExpert` and updates the manifest.
    - **Action 2 (Critical):** Executes **Surgical History Pruning**. The database record of the Concierge's refusal ("I currently lack...") is deleted.
3. **Phase 2: Seamless Execution**
    - The system prompts the user to continue (or processes the next turn).
    - **User Input:** "Did you send it?"
    - **System Context:** Because the refusal was deleted, the LLM context looks like: *User: "Send email..." -> System: [New Email Expert Activated]*.
    - **Response:** "I am ready. Please provide the recipient address."
    - **Result:** The agent successfully acquired the skill without the user needing to restart the session or manually install plugins.

## Scenario 3: Dead End (The "Flight" Test)

**Objective:** Validate fault tolerance against "Hollow Evolution" (hallucinating tools).

1. **Phase 1: Request**
    - **User Input:** "Book a flight to London for next Tuesday."
    - **System Action:** Concierge refuses: "I lack the capability to book flights."
2. **Evolution Check:**
    - Listener analyzes intent: `Book Flight`.
    - Registry Scan: Searches for `Flight_MCP` in Cold Storage.
    - **Result:** `None` (Tool does not exist in the repository).
3. **Outcome:**
    - **Action:** The system does **not** create a new agent.
    - **Reporting:** It generates a `GAP_REPORT` log entry: `[MISSING_FEATURE]: FlightBooking requested by UserID_001`.
    - **User Feedback:** "I have logged a request for flight booking capabilities with our engineering team."

- **Result:** Prevents the creation of a "dummy" agent that would simply apologize in a loop, thereby maintaining system trust.

## V. ASSUMPTIONS AND LIMITATIONS

### A. Assumptions

The research assumes a **stateless tool architecture** where MCP tools can be hot-swapped without complex configuration (e.g., OAuth tokens are pre-provisioned). It also assumes a **single-node runtime**; the current thread-pool implementation is designed for a single machine and would require refactoring for distributed clusters.

### B. Limitations

1. **Evolution Latency:** There is an inherent delay between the user's failed request and the system's evolution, determined by the Listener-Learner's polling interval.
2. **Cold Start Overhead:** Instantiating a new expert agent introduces a slight latency penalty compared to calling a pre-loaded tool.
3. **State Complexity:** Debugging an autonomously changing runtime is difficult, as reproducing bugs requires precise snapshots of the ephemeral registry state.

## VI. FUTURE WORK

Future iterations of this research will focus on the following areas to address current limitations and expand system capabilities:

1. **Distributed Containerization:** We aim to decouple experts from thread-based execution, moving them into isolated Docker containers or Kubernetes pods. This will enhance security by sandboxing code execution and allow for independent scaling of heavily used experts.
2. **Agent Snapshotting:** Currently, evicted experts lose their conversational context. We propose implementing a serialization mechanism to save the state of evicted agents to disk, allowing them to be "rehydrated" with their learned context intact.
3. **Dynamic Elasticity:** We plan to implement auto-scaling logic that dynamically adjusts the number of active expert slots (K) based on available system memory and real-time load, rather than relying on a fixed hard limit.

## VII. CONCLUSION

This work presents the Self-Evolving Concierge System, a robust architectural solution to the "Generalization-Specialization Dilemma" in autonomous AI agents. By synthesizing the flexibility of monolithic models with the precision of specialized swarms, the proposed Dynamic Mixture of Experts (DMoE) framework successfully mitigates context pollution without incurring the prohibitive computational costs of static multi-agent systems. The novel contributions of this research include the asynchronous "Listener-Learner" engine; the resource-efficient LRU eviction policy; and the "Surgical History Pruning" technique. These components collectively enable a runtime environment that is both adaptive and coherent. We have demonstrated that treating agent capabilities as ephemeral, managed resources rather than static assets allows for significantly higher scalability. As LLMs continue to integrate with an expanding universe of external tools, such "Just-in-Time" orchestration architectures will be foundational to building General Purpose AI that is both powerful and efficient.

## REFERENCES

[1] J. Fang et al., "A Comprehensive Survey of Self-Evolving AI Agents: A New Paradigm Bridging Foundation Models and Lifelong Agentic Systems," *arXiv preprint arXiv:2508.07407*, 2025.

[2] H. Gao et al., "A Survey of Self-Evolving Agents: On Path to Artificial Super Intelligence," *arXiv preprint arXiv:2507.21046*, 2025.

[3] N. F. Liu et al., "Lost in the Middle: How Language Models Use Long Contexts," *Transactions of the Association for Computational Linguistics*, vol. 12, pp. 1-15, 2024.

[4] C. S. Xia et al., "Live-SWE-agent: Can Software Engineering Agents Self-Evolve on the Fly?," *arXiv preprint arXiv:2511.13646*, 2025.

[5] N. Shazeer et al., "Outrageously Large Neural Networks: The Sparsely-Gated Mixture-of-Experts Layer," *ICLR*, 2017.

[6] Y. Cai et al., "Building Self-Evolving Agents via Experience-Driven Lifelong Learning: A Framework and Benchmark," *arXiv preprint arXiv:2508.19005*, 2025.

[7] J. S. Park et al., "Generative Agents: Interactive Simulacra of Human Behavior," *UIST '23: Proceedings of the 36th Annual ACM Symposium on User Interface Software and Technology*, 2023.

[8] D. Zhang, "AgentDevel: Reframing Self-Evolving LLM Agents as Release Engineering," *arXiv preprint arXiv:2601.04620*, 2026.

[9] Self Evolving Multi Agent Concierge System , https://www.kaggle.com/code/sathishsampathai/self-evolving-multi-agent-concierge-system